\documentclass[10pt, aps,prc,twocolumn,superscriptaddress,preprintnumbers,
amsmath, 
floatfix,
longbibliography,
nofootinbib
]{revtex4-1}
\usepackage[T1]{fontenc}
\usepackage{adjustbox}          % Used to adjust figure frames
\usepackage[caption=false]{subfig}
\usepackage{url}
\usepackage{color}
\usepackage{float}
\usepackage[pdftex,colorlinks=true, linkcolor = blue, citecolor=blue,urlcolor=blue, bookmarksnumbered=true, bookmarksopen=true]{hyperref}
\usepackage{longtable}
\usepackage{amsfonts}
\usepackage{dsfont}
\usepackage{wrapfig,bm} 
\usepackage[normalem]{ulem}
\usepackage{MnSymbol}
\usepackage{float}

\usepackage{lipsum}

\begin{document}
 
\title{ Quantal effect on the opening angle distribution between the fission fragment's spins }

\author{Guillaume Scamps}
\affiliation{Department of Physics, University of Washington, Seattle, WA 98195--1560, USA}
\affiliation{Laboratoire des 2 Infinis -Toulouse (L2IT-IN2P3), Université de Toulouse, CNRS, UPS, F-31062 Toulouse Cedex 9, France}
  \date{\today}

\begin{abstract}

\begin{description}
\item[Background] 
 Several approaches are currently trying to understand the generation of angular momentum in the fission fragments. The microscopic TDDFT and statistical FREYA lead to different predictions concerning the opening angle distribution formed between the two spins in particular at 0 and 180 degrees. 
 
\item[Purpose]  
This letter aims to investigate how the geometry and the quantum nature of spins impact the distribution of opening angles to understand what leads to different model predictions.
 
\item[Method] 
Various assumptions of K distribution (K=0, isotropic, isotropic with total K=0, and from TDFFT) are investigated in a quantum approach. These distributions are then compared to the classical limit using the Clebsch-Gordan coefficients in the limit of $\hbar$ approaches zero. 
 
\item[Results]  
It is shown that in all the schematic scenario the quantal distribution of opening angle lead to the expected behavior in the classical limit. The model shows that the quantal nature of the spins prevents the population of opening angles close to 0 and 180 degrees. The difference in opening angle in the 2D and isotropic 3D distribution is discussed and it is shown that the realistic TDFFT opening angle distribution presents an intermediate behavior between the two cases.

\item[Conclusions] 
The last comparison reveals two key differences between the two model's predictions: the quantal spins' nature in TDDFT and the assumption of zero K values in FREYA.

\end{description}

\end{abstract}

\maketitle   

 \vspace{0.5cm}
 
 Nuclear fission, the process in which the nucleus of an atom splits into two smaller fragments, is a fundamental phenomenon in nuclear physics with far-reaching implications, from nuclear energy generation to our understanding of the universe's evolution. While the basics of fission are well-known \cite{Andreyev_2018,Schmidt:2018,SCHUNCK2022103963}, there is an intriguing aspect that has recently captured the attention of the theoretical community, the determination of the spin opening angle \cite{Bender_2020}.  This angle represents the angle between the angular momentum (or spin) of the two fission fragments. It holds valuable insight into the dynamics of the fission process since the distribution of this opening angle is shaped by the geometrical characteristics of the spin distributions of these fragments and their mutual correlation modes: tilting, twisting, wriggling, or bending modes.

\begin{figure}[h]
\centering
\includegraphics[width=0.95\columnwidth]{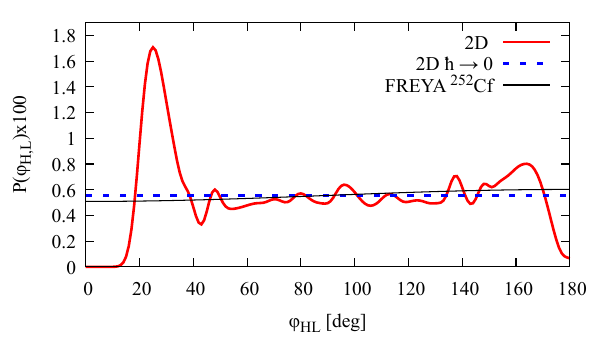}
\caption{ Opening angle distribution assuming a 2D geometry and a spin cut-off distribution. The distribution in the classical limit is also compared to  FREYA \cite{Vogt:2021}. }
\label{fig:2D}
\end{figure}

\begin{figure}[t]
\centering
\includegraphics[width=0.95\columnwidth]{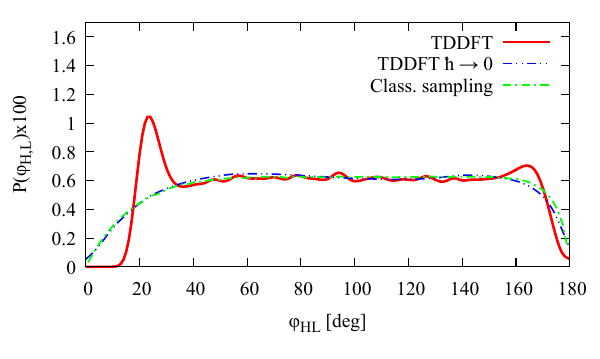}
\caption{ Opening angle distribution for the $^{252}$Cf spontaneous fission obtained with a TDDFT calculation \cite{Scamps:2023a} as well as its corresponding distribution in the classical limit. The green curve shows the result of a sampling statistical distribution with the same $\theta_F$ angle distribution as in the TDDFT calculation. }
\label{fig:TDDFT}
\end{figure}

The opening angle has been introduced in Ref.~\cite{Randrup:2021,Vogt:2021}  in the framework of the FREYA model. This model uses a classical approximation and predicts an essentially flat distribution between 0 to 180 degrees (Shown in Fig. \ref{fig:2D}). This was followed by a microscopic calculation~\cite{Bulgac:2022b} predicting a completely different distribution with vanishing probabilities at 0 and 180 degrees and a strong asymmetry suggesting a large population of the bending mode (the spins are mostly in opposite directions). However, the use of a simplified projection operator led to a significant inaccuracy in that distribution. That was corrected in Ref.~\cite{Scamps:2023a} with an exact projection operator in the same Time-dependent density function theory (TDDFT) framework. The resulting distribution (shown in Fig.~\ref{fig:TDDFT}) does only slightly favor the bending mode and is mostly flat, but still, with a vanishing population at 0 and 180 degrees. A similar result is obtained with the collective Hamiltonian model \cite{Scamps:2023}. In the recent TD-MRV model \cite{Chen23} the open angle distribution also follows that type of distribution with an additional peak at 100 degrees.

While experimental data suggests that spin magnitudes are not significantly correlated \cite{Wilson:2021}, the determination of the opening angle is a more challenging task \cite{Randrup:2022a}. Consequently, in the near future, we will likely depend on theoretical analysis.
In Ref. \cite{Randrup:2022},  an interesting discussion is presented about the impact of the geometry (2D or 3D) on the opening angle distribution in a classical framework. In the present contribution, I propose a similar analysis in a quantum framework, to understand how the quantum nature of the spins impacts the classical picture and to understand the reasons of the discrepancies between the models. 

 Disregarding the mechanism leading to the spin of the fragments, when the fragments are separated by a large distance, the final state can be written in the form, 
 \begin{align}
 |\Psi\rangle = \sum_{S_H,K_H,S_L,K_L} c_{S_H,K_H,S_L,K_L} |S_H,K_H,S_L,K_L \rangle,
 \end{align}
 describing the heavy (H) and light (L) fragment's spin ($S_{H,L}$) and projection ($K_{H,L}$)  on the z-axis (the fission axis) with $-S_F \leqslant K_F \leqslant S_F$. It has to be noted that the K quantum number here is not the projection of the spin in the intrinsic framework of each fragment but on the fission axis. In this letter, I investigate several cases of uncorrelated limits, assuming different geometries.
 \begin{itemize}
 \item 2D : Both K are zero, the spins of the fragments are perpendicular to the fission axis
 \item Isotropic 3D : The orientation of both spins are fully isotropic corresponding to a uniform and uncorrelated distribution of $K_H$, $K_L$.
 \item  Isotropic 3D : with total $K=0$ : The spins are constrained with $K_H=-K_L$ which is necessary to ensure that the total spin of the system is zero.
 \end{itemize}
 
 In all cases, the total spin of the system is, 
 \begin{align}
 \mathbf{S_H} +  \mathbf{S_L}  +  \mathbf{\Lambda}  =  \mathbf{S_0}. 
 \end{align}
 Although previous contributions focus on the simpler case with a zero total spin, considering the vector $\mathbf{\Lambda'}  =   \mathbf{\Lambda} - \mathbf{S_0}$, the same triangle rule can be established between the three vectors  $\mathbf{S_H}$, $\mathbf{S_L}$,  and  $\mathbf{\Lambda'}$. The difference is that $\Lambda'_z$ the projection of $\mathbf{\Lambda'}$ on the fission axis can be different than 0.

 The distribution of the opening angle is then given by, 
  \begin{align}
  &P( \varphi_{HL} ) = \nonumber \\ 
  & \sum_{\substack{\Lambda', \Lambda'_z, \\ S_H \ne 0 , S_L \ne 0}}
 \delta \left( \varphi_{HL}-  \varphi_{HL}(  \Lambda', S_H, S_L) \right)    P(  \Lambda', \Lambda'_z , S_H, S_L), \\
 & \varphi_{HL} = \arccos\left( \frac{ \Lambda' (\Lambda' +1) - S_H (S_H +1) - S_L(S_L+1) }{2 \sqrt{S_H(S_H + 1) S_L(S_L + 1)}  } \right), \\
& P(  \Lambda', \Lambda'_z , S_H, S_L)  = \nonumber \\ 
 & \quad  \quad \quad  \quad  \left| \sum_{K_H,K_L}  C_{S_H, K_H,S_L, K_L}^{\Lambda',\Lambda'_z}   c_{S_H,K_H,S_L,K_L}   \right|^2.
 \end{align}
 Note that this distribution is always discrete with a lot of contributions, in practice to visualize the overall distribution shape, each peak is folded with a Gaussian of width 3 degrees.
 
 Starting with the 2D case, in which both spins are perpendicular to the fission axis, i.e. only the $K_F$=0 states are populated, and assuming a spin cutoff distribution for the spin, 
 \begin{align}
 |c_{S_H,K_H,S_L,K_L} |^2 \propto & \delta_{K_H,0} \delta_{K_L,0} (2 S_H + 1 ) e^{\frac{-S_H(S_H+1)}{2\sigma_H^2}} \nonumber \\ 
 & \times  (2 S_L + 1 ) e^{\frac{-S_L(S_L+1)}{2\sigma_L^2}}.
 \end{align}
 Where the magnitudes of the spins are uncorrelated, which is consistent with both the experimental data \cite{Wilson:2021} and theory \cite{Bulgac:2022b,Scamps:2023,Randrup:2022}.
 A phase in the coefficient would not change the resulting distribution shown in Fig.~\ref{fig:2D}. The parameters of the spin cut-off distribution are chosen to reproduce  the microscopic results \cite{Bulgac:2021} with $\sigma_H=7 \hbar$ and $\sigma_L=10 \hbar$.   The quantal 2D distribution is not uniform as it shows a peak at 25 degrees and a depletion at 0 and 180 degrees. To better understand the depletion at 0 degrees, it can be determined, in a spin distribution limited by  $S_H$ and $S_L < S_c$, the combination of spin $(S_H, S_L, \Lambda')$ leading to the minimal angle. This one is found for $S_H=S_L=S_c$ and $\Lambda'=S_L+S_H$ and lead to a minimal angle $\varphi_{HL}^{min}=\arccos(\frac{S_c}{S_c+1})$. This function is very slowly decaying to zero.  Obtaining an opening angle of 5 degrees demands a value of $S_c$ at 262 $\hbar$, and for a 1-degree angle, an exceedingly large value of 6565 $\hbar$, both are excessively unrealistic in the present context.

To determine the classical equivalent of that distribution, the limit $\hbar\rightarrow 0$ can be taken using the classical limit of the Clebsch-Gordan coefficients~\cite{BRUSSAARD:1957,Varshalovich:1988}. The open angle distribution becomes, 
\begin{widetext}
\begin{align}
P( \varphi_{HL} ) =& \sum_{\substack{K_H, K_L, \\ S_H, S_L}} |c_{S_H,K_H,S_L,K_L}|^2 
(C_{S_H, K_H,S_L, K_L}^{{\rm class.   }\Lambda',\Lambda'_z})^2 \sin(\varphi_{HL}) \frac{S_H S_L}{\Lambda} , \\
 (C_{S_H, K_H,S_L, K_L}^{{\rm class.   }\Lambda',\Lambda'_z})^2 = & \frac{2 \Lambda' }{\pi} \left( -\left( S_H^4 + S_L^4 + \Lambda'^4 \right) + 2 \left( S_H^2 S_L^2 + S_H^2 \Lambda^2 + S_L^2 \Lambda^2 \right) + 4 \left( - S_H^2 K_L \Lambda'_z - S_L^2 K_H \Lambda'_z  + \Lambda^2 K_H K_L  \right)   \frac{}{}  \right)^{-1/2}.
\end{align}
\end{widetext}

  \begin{figure}[h]
\centering
\includegraphics[width=0.95\columnwidth]{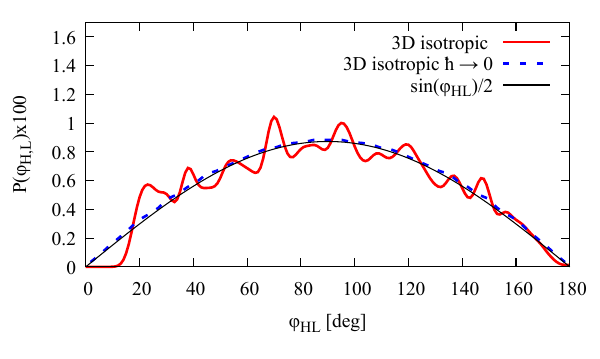}
\includegraphics[width=0.95\columnwidth]{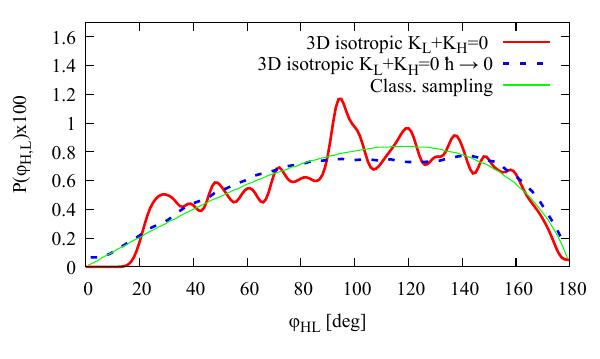}
\caption{ Top: Opening angle distribution assuming a 3D isotropic geometry and a spin cut-off distribution. The distribution in the classical limit is also shown with a dashed blue line. Bottom: Same with the constraint that the z component of the spins are in opposite directions. The green line shows the result of a 3D sampling.}
\label{fig:3Duniform}
\end{figure}

 In the case of a 2D distribution, the distribution is found to be uniform, 
 \begin{align}
%P(\varphi_{H,L}) &=  \frac{ c_{S_L,S_H}^2  (C^{{\rm class.   } \Lambda 0}_{S_H,0,_L,0})^2 \sin( \varphi_{HL} )  {S_H S_L} }{ \Lambda}\\
 %P(\varphi_{HL} ) &= \sum_{S_L,S_H} \frac{c_{S_L,S_H}^2   (2/\pi) \sin( \varphi_{HL} )  {S_H S_L} } {\sqrt{  -( S_H^4 + S_L^4 + \Lambda^4 ) + 2(S_H^2S_L^2 + S_H^2\Lambda^2+ S_L^2\Lambda^2)  } } \\
P(\varphi_{HL} ) &= \sum_{S_L,S_H} c_{S_L,S_H}^2 / \pi, 
\end{align}
which is close to the FREYA distribution, as seen in Fig.~\ref{fig:2D}. The uniform distribution is the one that is expected in classical physics for two independent spins in the same plane. Although classical physics allows two spins to be correlated in direction and confined in a 2-dimensional plane, it is not allowed in nuclear physics where spatial correlations require the population of non-zero K.

 Another hypothesis is a fully isotropic distribution assuming all the K of each fragment to be equiprobable. In that hypothesis, the fragments do not respect the $K_H+K_L=0$ rule that is mandatory in the S=0 case.
The coefficients of the two-spin wave packet are assumed to be 
 \begin{align}
 |c_{S_H,K_H,S_L,K_L} |^2 \propto &    e^{\frac{-S_H(S_H+1)}{2\sigma_H^2}}      e^{\frac{-S_L(S_L+1)}{2\sigma_L^2}}.
 \end{align}
 The phase of each coefficient is chosen randomly. The fragments are then completely independent. In that situation, in the classical limit, the opening angle is expected to follow a distribution $P(\varphi_{HL})=\sin(\varphi_{HL}))/2$. The quantal distribution shown in Fig.~\ref{fig:3Duniform} follow roughly this behavior with fluctuation around that curve due to its quantum nature. Employing the classical limit of the Clebsch-Gordon coefficients results in the same distribution than in the classical sampling.

 The previous case represents an ideal scenario where the total spin is unconstrained. In the case of $S_0=0$ like in the spontaneous fission of $^{252}$Cf, an additional condition has to be taken into account, 
 \begin{align}
& |c_{S_H,K_H,S_L,K_L} |^2 \propto & \nonumber \\
   &\quad \frac{ \delta_{K_H-K_L}   (2 S_H + 1 ) e^{\frac{-S_H(S_H+1)}{2\sigma_H^2}}   (2 S_L + 1 ) e^{\frac{-S_L(S_L+1)}{2\sigma_L^2}} }{ 2 \min(S_H,S_L)+1},
 \end{align}
 with the denominator a normalization factor to ensure the same distribution $P(S_H,S_L)$ as in the two previous cases.
 The three resulting distributions are shifted to the right. Indeed, the opening angle is enlarged because the spins are required to have opposite z-components.  This last scenario is equivalent to the one of Ref. \cite{Bulgac:2022e} assuming an uniform distribution of $\Lambda$ constrained to respect the triangular rule.
 
The previous examples showed that the 2D geometry leads to a flat gauge angle distribution while the fully isotropic 3D gives a distribution close to the sinus function. To finish, I consider the case of the realistic TDDFT calculation  \cite{Scamps:2023a} which is obtained using the projection method \cite{Simenel:2010,Scamps:2013,Scamps:2015}. The microscopic model shows an intermediate behavior with a $K$ distribution decaying quickly as a function of $K$. Equivalently, the angle between the spins and the fission axis $\theta_F$ is contained in a distribution at an angle close to 90 degrees \cite{Scamps:2023a}. This leads to an intermediate behavior between the 2D and isotropic 3D scenario as seen in Fig. \ref{fig:TDDFT}. The distribution is mostly flat but decays for angles close to 0 and 180 degrees. 

The distribution in the limit of $\hbar \rightarrow 0 $ can also be obtained with a classic sampling of the two spins. This sampling assumes the same spin cutoff distribution for the magnitude of the spins as previously, the orientation of the spins are however constrained with a distribution of angle $\theta_H$,
\begin{align}
P(\theta_H)  \propto e^{ - \frac{ |\theta_H-90| } {0.28} },
\end{align}
and with a $K_H=-K_L$ to determine the other polar angle. The distribution $P(\theta_H)$ is chosen to be close to the TDDFT results for  $^{252}$Cf~\cite{Scamps:2023a}. This sampling of the spins produces a distribution of opening angle that reproduces the TDDFT distribution in the case where the Clebsch-Gordon coefficients are replaced by their classical equivalent.

 From that last comparison, it is possible to understand the main difference between the TDDFT and FREYA opening angle distribution. This difference can be attributed to two primary factors. Firstly, the quantal nature of the spins in TDDFT prevents the population of opening angle close to 0 degrees and depletes the region around 180 degrees.  The second reason is the population of non-zero K, which also reduces the probability distribution of the two regions at the edge of the interval.
 
  \vspace{0.5cm}

{\bf Acknowledgements} \\

I thank Aurel Bulgac and J\o{}rgen Randrup for interesting discussions. 
The funding from the US DOE, Office of Science, Grant No. DE-FG02-97ER41014 is greatly appreciated.

%%%%%%%%%%%%%%%%%%%%%%%%%%%%%%%%%%%%%%%%%%%%%

% These are needed to avoid a babel error.
\providecommand{\selectlanguage}[1]{}
\renewcommand{\selectlanguage}[1]{}

\bibliography{local_fission.bib}

\end{document}